\documentclass[aps,reprint,superscriptaddress,preprintnumbers,amsmath,amssymb,dvipdfmx]{revtex4-2}
\usepackage{dcolumn}% Align table columns on decimal point
\usepackage{bm}% bold math
\usepackage[utf8]{inputenc}
\usepackage[T1]{fontenc}
\usepackage{mathptmx}
\usepackage{etoolbox}
\usepackage[dvipdfmx]{graphicx}
\usepackage{color}
\usepackage{amsmath}
\usepackage{amsfonts}
\usepackage{amssymb}
\usepackage{comment}
\usepackage{siunitx}
\usepackage{here}

\begin{document}
% \linenumbers

\newcommand{\kymemo}[1]{{\color{magenta}(#1)}}

\preprint{APS/123-QED}

\title{%Dynamics of lane formation of particles immersed in viscoelastic fluids
Dynamics of particle lane formation in confined viscoelastic fluids under shear}% Force line breaks with \\

\author{Hiroto Yokoyama}%
 \affiliation{JGC Corporation, Process Technology Division, EN Technology Center, 2-3-1 Minatomirai, Nishi-ku, Yokohama-shi, Kanagawa, 220-6001, Japan}%
\author{Masanori Honda}
 \affiliation{Department of Applied Physics, Tokyo University of Science, 6-3-1 Nijuku, Katsushika-ku, Tokyo, 125-8585, Japan}
\author{Rinya Miyakawa}
 \affiliation{Department of Applied Physics, Tokyo University of Science, 6-3-1 Nijuku, Katsushika-ku, Tokyo, 125-8585, Japan}
 \author{Yuki Shinohara}
 \affiliation{Department of Applied Physics, Tokyo University of Science, 6-3-1 Nijuku, Katsushika-ku, Tokyo, 125-8585, Japan}%
 \author{Kota Nakamura}%
 \affiliation{Department of Applied Physics, Tokyo University of Science, 6-3-1 Nijuku, Katsushika-ku, Tokyo, 125-8585, Japan}%
 \author{Kojiro Otoguro}%
 \affiliation{Meiji Institute for Advanced Study of Mathematical Science (MIMS), Meiji University, 4-21-1, Nakano, Nakano-ku, Tokyo 164-8525, Japan}%
\author{Kiwamu Yoshii}%
 \affiliation{Department of Applied Physics, Tokyo University of Science, 6-3-1 Nijuku, Katsushika-ku, Tokyo, 125-8585, Japan}%
\author{Yutaka Sumino}\email{ysumino@rs.tus.ac.jp}
\affiliation{Department of Applied Physics, Tokyo University of Science, 6-3-1 Nijuku, Katsushika-ku, Tokyo, 125-8585, Japan}%
\affiliation{
 Water Frontier Science \& Technology Research Center, and Division of Colloid Interface, Research Institute for Science \& Technology, Tokyo University of Science, 6-3-1 Nijuku, Katsushika-ku, Tokyo, 125-8585, Japan
}%
\affiliation{
 Faculty of Engineering and Physical Sciences, University of Surrey, Guildford, Surrey GU2 7XH, United Kingdom
}%

\date{\today}% It is always \today, today,

\begin{abstract}
Simple shear flow can induce flow-aligned chain formation of particles suspended in viscoelastic fluids. Although this phenomenon has been reported for decades, direct {\it in situ} measurements of the alignment dynamics and particle trajectories during chain formation remain limited. Here, we develop an {\it in situ} observation platform based on parallel rotating disks separated by a gap comparable to the particle diameter, enabling simultaneous observation of particle alignment under radially varying shear rates. The narrow gap strongly confines particle motion, thereby enhancing hydrodynamic interactions and collision events between particles. Using a viscoelastic fluid embedding zircon particles as the sample, we find that alignment occurs once the local particle Weissenberg number exceeds unity (Wi$_\mathrm{p} \geq 1$), defined using an effective shear rate based on the wall velocity and the available gap width. Particle tracking further reveals a back-and-forth shuttling motion that accompanies the alignment process. Using the image brightness in a colored fluid as a proxy for out-of-plane position, we show that the shuttling originates from vertical displacement of the particles. We further construct a minimal agent-based model in which the vertical particle position follows a Ginzburg–Landau-type double-well potential, and demonstrate that collision-driven accumulation emerges in numerical simulations. In the strongly confined geometry, alignment occurs by an effective attraction due to collision, which is reminiscent of motility-induced clustering often observed in active matter.
\end{abstract}

\maketitle

%\tableofcontents

\section{Introduction}
From the viewpoint of continuum mechanics, Newtonian fluids generate stress proportional to the applied strain rate, whereas many complex fluids, including polymeric liquids, exhibit non-Newtonian behavior~\cite{Bird1987}.
Such fluids possess internal microstructures that give rise to memory effects and elastic responses under deformation, enabling the emergence of dynamic structures even under simple shear flows.
Examples of flow-induced structures in viscoelastic fluids are found across a wide range of systems in nature and industry~\cite{Datta2022-pr}.

A particularly well-known example of such non-Newtonian behavior is the spontaneous alignment of particles dispersed in viscoelastic fluids under shear~\cite{Michele1977,Won2004,Scirocco2004,Pasquino2010,Pasquino2010a,Pasquino2013,Pasquino2014,VanLoon2014}.
Particles initially distributed randomly between parallel plates can self-organize into flow-aligned structures, often referred to as lane formation.
Similar particle and bubble alignment phenomena play important roles in microfluidic manipulation and separation~\cite{Li2020a}, industrial multiphase flows~\cite{Barros2022}, and potentially in geological~\cite{Wallis2021,TIAN2021} and biological~\cite{Napolitano2007,Douezan2012} systems.

Previous studies have identified the first normal stress difference as a key mechanism driving particle alignment in viscoelastic shear flows~\cite{Michele1977}.
Shear-thinning effects~\cite{VanLoon2014}, as well as confinement by solid boundaries~\cite{Choi2012}, have also been suggested to enhance lane formation, and numerical studies support these interpretations~\cite{Jaensson2016}.
However, despite extensive investigations of the aligned structures themselves, comparatively little attention has been paid to the dynamics of individual-particle motion that lead to alignment.
Understanding particle-scale dynamics is essential for controlling alignment phenomena in applications such as particle separation, accumulation, and transport.

In this study, we address this gap by developing an experimental system in which particles dispersed in a viscoelastic fluid are confined between a transparent acrylic upper plate and a bottom disk, arranged in a parallel configuration with a gap comparable to the particle diameter.
The bottom disk is rotated unidirectionally at a constant angular velocity, producing a confined shear flow with a shear rate that varies systematically with radial position.
The lower disk is smaller than the upper plate, yet its diameter is still more than two orders of magnitude larger than the particle diameter, allowing us to approximate locally unidirectional shear while simultaneously accessing a wide range of shear rates in a single experiment.
Compared with reciprocating parallel-plate geometries used in earlier studies~\cite{Michele1977}, this setup enables continuous shear over long durations, facilitating direct observation of transient and steady-state particle dynamics.

Using this platform, we track particle motion from an initially disordered state to the emergence of aligned structures.
We show that particle alignment occurs when the local Weissenberg number near the particles ($\rm Wi_p$) exceeds unity.
In the present confined geometry, the relevant control parameter is not the Weissenberg number defined by the plate separation, but rather a local Weissenberg number defined in the vicinity of a particle.
Because the particle occupies a significant fraction of the gap, the effective fluid thickness is reduced, leading to an enhanced elastic response near the particle.
We therefore introduce a particle-based Weissenberg number, $\mathrm{Wi_p}$, which quantifies the local balance between elastic and viscous effects around an individual particle.
In the same regime, we uncover a characteristic shuttling motion in the flow direction, which is absent at lower Weissenberg numbers.
Single-particle analysis reveals that this shuttling originates from bistability in the particle height induced by viscoelastic normal stresses.
Finally, we introduce a minimal numerical model that couples height bistability with excluded-volume interactions and demonstrate that this mechanism reproduces the observed alignment through an effective attraction between particles.

\section{Experimental Methods}

%====================================
%====================================
\begin{figure}[htb]
\centering
\includegraphics{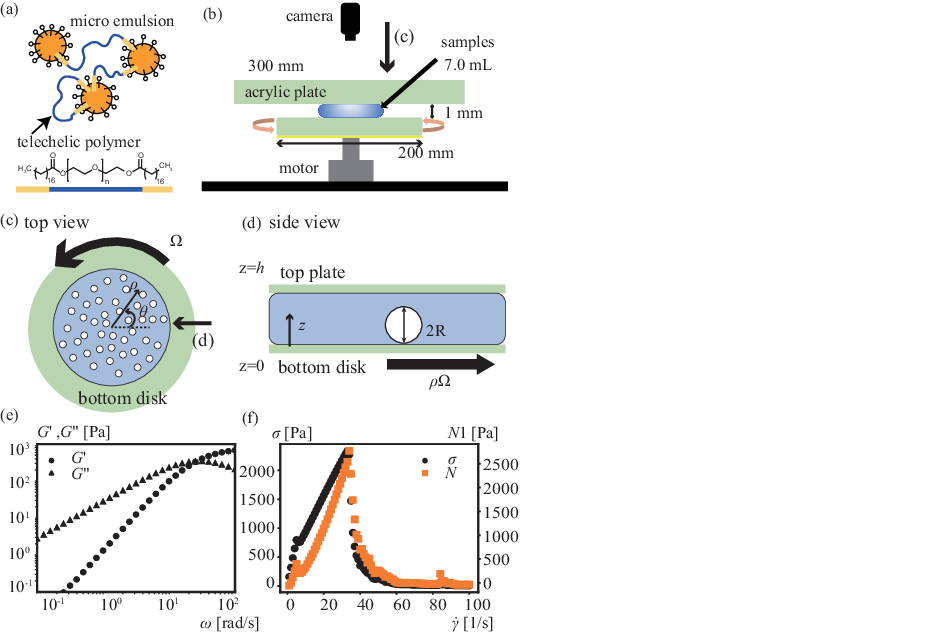}
\caption{\label{fig:figure_1} 
(a) Schematic illustration of the sample: oil-in-water microemulsions bridged by telechelic polymers.
(b) Schematic illustration of the experimental setup, showing the parallel-plate configuration with a rotating bottom disk and optical access for in situ observation.
(c,d) Definition of the observation geometry and coordinate systems.
(c) Top view of the bottom disk, defining the radial ($\rho$) and azimuthal ($\theta$) directions.
(d) Cross-sectional view around a particle confined between the plates, defining the height ($z$) direction and the effective gap $h-2R$.
(e) Storage and loss moduli, $G'$ and $G''$, obtained from small-amplitude oscillatory shear.
(f) Steady shear stress, $\sigma$, and the first normal stress difference, $N_1$, as functions of shear rate $\dot{\gamma}$.}
\end{figure}
%====================================
%====================================

\begin{figure*}[tb]
\includegraphics{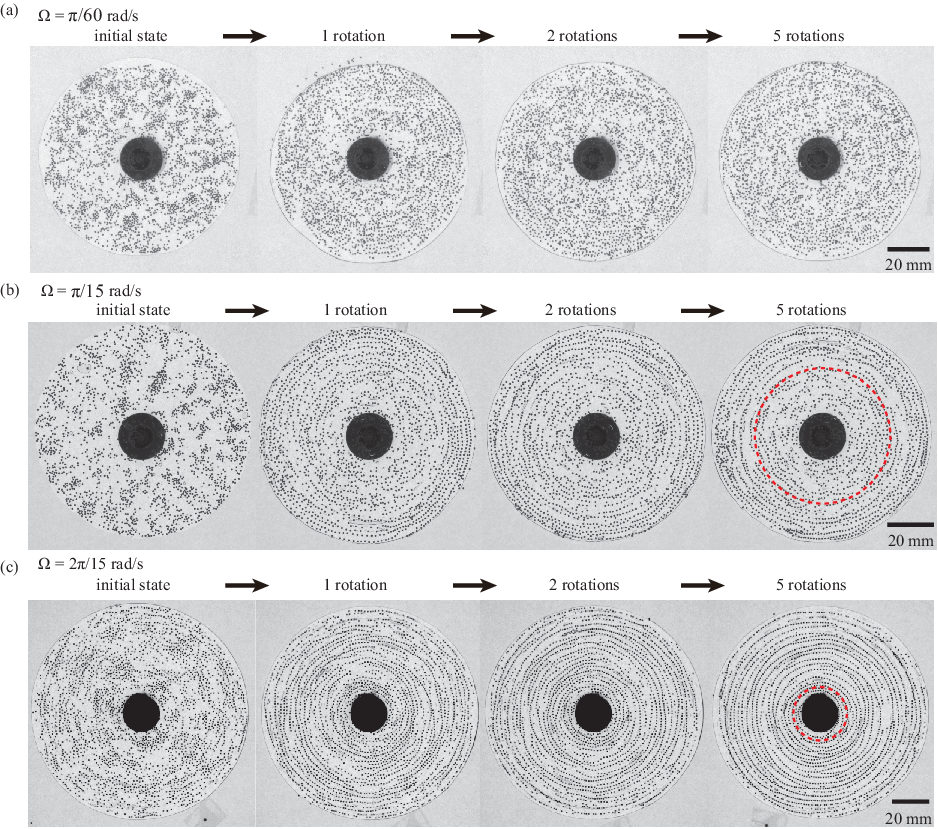}
\caption{\label{fig:figure_2}
Snapshots of the particle suspension at three angular velocities, shown at four values of the accumulated rotation angle $\Theta=\Omega t$ ($0,\,2\pi,\,4\pi,$ and $10\pi$).
(a) $\Omega=\pi/60$~\SI{}{\radian\per\second};
(b) $\Omega=\pi/15$~\SI{}{\radian\per\second};
(c) $\Omega=2\pi/15$~\SI{}{\radian\per\second}.
The red dashed circle indicates the critical radius $\rho_c$ at which the local Weissenberg number near a particle satisfies $\mathrm{Wi_p}=1$, where $\rho_c=(h-2R)/(\lambda\Omega)$.
Lane formation is observed in regions where $\mathrm{Wi_p}>1$. 
Scale bar: \SI{20}{\milli\meter}.
}

\end{figure*}
% ====================================
% ====================================

\begin{figure}[tb]
\includegraphics{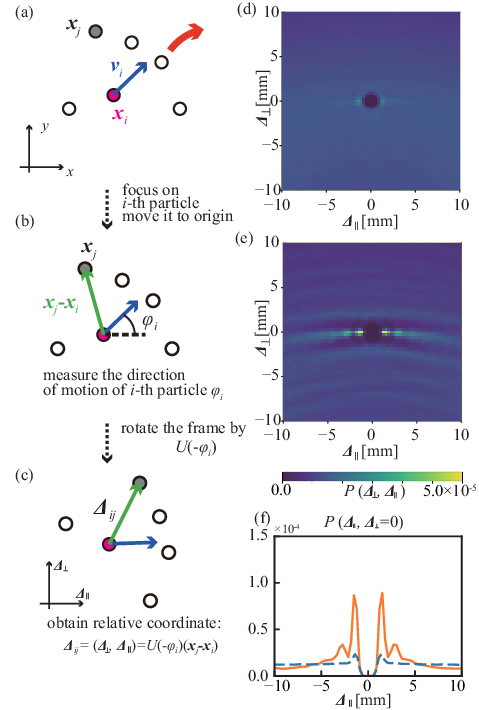}
\caption{\label{fig:figure_3}
Autocorrelation analysis of particle positions in a reference frame centered on a particle and aligned with its instantaneous velocity.
(a--c) Schematic illustration of the coordinate transformation used to fix the direction of motion of a reference particle.
Particle positions are expressed in a frame rotated such that the velocity of the reference particle defines the $\Delta_\parallel$ direction.
(d,e) Spatial autocorrelation function of particle positions, $P(\Delta_\parallel,\Delta_\perp)$, obtained at $\Omega=\pi/60$~\SI{}{\radian\per\second} (d) and $\Omega=2\pi/15$~\SI{}{\radian\per\second} (e).
The horizontal and vertical axes correspond to the directions parallel ($\Delta_\parallel$) and perpendicular ($\Delta_\perp$) to the particle motion, respectively.
(f) Cross sections of $P(\Delta_\parallel,\Delta_\perp)$ at $\Delta_\perp=0$, corresponding to panels (d) blue dashed line and (e) orange solid line.
}
\end{figure}
% %====================================
% %====================================
\subsection{Samples}
We used spherical zircon beads (ZS0608, Hira Ceramics Co., Ltd.) with radius $R=\SI{0.35}{\milli\meter}$ and density $4.4\times10^{3}\,\SI{}{\kilo\gram\per\cubic\meter}$ as dispersed particles. We prepared a viscoelastic fluid consisting of an oil-in-water microemulsion bridged by telechelic polymers~\cite{Filali1999b,Michel2001,Tabuteau2009,Tabuteau2011,Tixier2010a}, which is schematically illustrated in Fig.~\ref{fig:figure_1}(a). Sodium chloride, octanol, and decane were purchased from Fujifilm Wako Pure Chemical Co., Ltd.; cetylpyridinium chloride (CpCl) from Tokyo Chemical Industry Co., Ltd.; and poly(ethylene glycol) distearate 6000 (telechelic polymer) from Polysciences, Inc. (Catalog No.~19234-100). Pure water was purified using a Millipore Milli-Q system.

We first prepared a microemulsion at approximately 20~wt.\% by mixing octanol, decane, CpCl, and an aqueous \SI{0.2}{\mole\per\liter} NaCl solution in a weight ratio of 1:3.1:4:32.4~\cite{Filali1999b}. The emulsion was agitated for \SI{48}{\hour} with a magnetic stirrer. Assuming a droplet diameter of \SI{14}{\nano\meter}, the droplet volume fraction was 25\%. The telechelic polymer and the emulsion were then mixed at a mass ratio of 0.0113:1 at room temperature and stirred for \SI{48}{\hour} to ensure complete dissolution. The sample was subsequently allowed to rest for \SI{24}{\hour} to remove entrained air bubbles. Under these conditions, the network connectivity, defined as the average number of hydrophobic alkyl chains per droplet, was approximately 9. For visualization purposes, we added 0.02~wt.\% Brilliant Blue FCF (Fujifilm Wako Pure Chemical Co., Ltd.) to the viscoelastic fluid.

\subsection{Observation setup}
Figures~\ref{fig:figure_1}(b) and (c) show the experimental setup, which consists of two parallel acrylic plates: an upper square plate and a lower circular disk.
The upper plate has dimensions of \SI{300}{\milli\meter}$\times$\SI{300}{\milli\meter}, and the bottom disk has a radius of \SI{100}{\milli\meter}.
Both the plate and the disk have a thickness of \SI{10}{\milli\meter}.
The center of the bottom disk is coupled to a rotating motor (BLM230-GFV2 equipped with GFV2G200 and BMUD30-A2; Oriental Motor Co., Ltd.), and the rotational speed is controlled by the BMUD30-A2 controller.

We defined a Cartesian coordinate system $(x,y,z)$ with the origin at the center of the bottom plate; the $z$-axis was taken along the plate gap [Fig.~\ref{fig:figure_1}(d)]. Because only in-plane motion is relevant here, the position and two-dimensional velocity of the $i$-th particle at time $t$ were expressed as $\bm{r}_i(t)=(x_i(t),y_i(t))$ and $\bm{v}_i(t)=(v^x_i(t),v^y_i(t))$, respectively. In addition, we defined a cylindrical coordinate system centered at the plate origin. The position of the $i$-th particle is expressed as $\bm{r}_i=\rho_i\,\bm{e}_{\rho,i}$, where $\rho_i=\sqrt{x_i^2+y_i^2}$, $\bm{e}_{\rho,i}=(\cos\theta_i,\sin\theta_i)$, and $\theta_i=\arctan\!\big(y_i/x_i\big)$  [Fig.~\ref{fig:figure_1}(c)]. The azimuthal unit vector was $\bm{e}_{\theta,i}=(-\sin\theta_i,\cos\theta_i)=U\!\left(\frac{\pi}{2}\right)\bm{e}_{\rho,i}$, and the positive $\theta_i$ direction was chosen to coincide with the rotational direction of the bottom plate. In this cylindrical coordinate system, the two-dimensional velocity of the $i$-th particle at time $t$ is expressed as $\bm v_i(t) =  v^{\rho}_i(t)\bm{e}_{\rho, i}+ v^{\theta}_{i}(t)\bm{e}_{\theta, i}$, and $d\theta_i/dt=\omega_i=v^{\theta}_{i}(t)/\rho_i(t)$

A volume of viscoelastic fluid was \SI{6}{ml}, and the particles with a total mass of \SI{2.2}{g} (mass density: $4.4 \times 10^3$ \SI{}{\kilo \gram \per \cubic \meter}) were dispersed in the viscoelastic fluid. Approximately 7.7\% of the sample in volume was filled with particles.
A viscoelastic fluid was placed on the bottom plate, and zircon beads were placed on top of the fluid.
Then, we left a sample still for 5 minutes to allow the beads to reach the bottom surface. The top plate was then situated to have the gap width $h$=\SI{1.0}{\milli \meter}. The bottom plate was rotated with an angular velocity $\Omega$.
We recorded images using a CMOS camera (DMK37BUX273) purchased from Imaging Source at frame rates ranging from 5 to 120 fps, depending on the rotation speed $\Omega$.

\subsection{Rheology of samples}

Rheological measurements were performed on an Anton Paar MCR~302 rheometer equipped with a cone--plate geometry (CPL-25) at \SI{25}{\celsius}. We conducted small-amplitude oscillatory shear (SAOS) tests and steady-shear measurements of the shear stress $\sigma$ and the first normal stress difference $N_{1}$ on the base viscoelastic fluid \emph{without} zircon beads.

Figure~\ref{fig:figure_1}(e) shows the SAOS response obtained at a strain amplitude $\Delta \gamma= 10$\%. The storage and loss moduli, $G'$ and $G''$, are well described by a single-mode Maxwell model,
\begin{align}
     G'(\omega) = \frac{G\,\lambda^{2}\omega^{2}}{1+\lambda^{2}\omega^{2}},
  \qquad
  G''(\omega) = \frac{G\,\lambda\,\omega}{1+\lambda^{2}\omega^{2}},
\end{align}
yielding a modulus $G \approx \SI{6.9e2}{\pascal}$ and a relaxation time $\lambda \approx \SI{4.9e-2}{\second}$. The corresponding zero-shear viscosity is $\eta_{0} = G\lambda \approx \SI{3.3e1}{\pascal\second}$.
%==
Figure~\ref{fig:figure_1}(f) summarizes the steady-shear response. The shear rate $\dot{\gamma}$ was increased in stepwise increments; each step was held for \SI{10}{\second}, and the average over the latter half of each step was reported. For $\dot{\gamma}$ < $\SI{33}{\per\second}$, the shear stress scales linearly with shear rate ($\sigma \propto \dot{\gamma}$), and the sample behaves as a viscous liquid with viscosity consistent with $\eta_{0}$. For $\dot{\gamma}$ > $\SI{33}{\per\second}$, the measured shear stress drops sharply, indicating a transition to an elastic/brittle response and sample failure~\cite{Filali1999b}. The first normal stress difference $N_{1}$ was recorded concurrently during these steady-shear tests.
For completeness, definitions of variables and coordinate systems are summarized in Appendix~\ref{app:variables}.

\section{Results and discussions}\label{sec:result}
\subsection{Experiment with a large number of particles: Lane formation}

By applying shear using a parallel rotating-plate geometry, we observed the collective rearrangement of particles dispersed in a viscoelastic fluid. Representative snapshots of the particle configurations are shown in Fig.~\ref{fig:figure_2}. Here, $\Theta$ denotes the accumulated rotation angle of the bottom plate, defined as $\Theta=\Omega t$, where $t=0$ corresponds to the onset of rotation. At $\Theta=0$, particles are randomly distributed in all cases.

For a low angular velocity, $\Omega=\pi/60$~\SI{}{\radian\per\second}, the particle distribution remains disordered even up to $\Theta=10\pi$~\SI{}{\radian} (five full rotations) [Fig.~\ref{fig:figure_2}(a)].
In contrast, for a higher angular velocity, $\Omega=2\pi/15$~\SI{}{\radian\per\second}, particles gradually align and form elongated structures parallel to the plate motion, i.e., along the $\theta$ direction [Fig.~\ref{fig:figure_2}(c)].
Here, we refer to the local dense aggregates aligned in the azimuthal direction as chain-like structures; their spatial arrangement in the radial direction gives rise to a macroscopic lane structure, and we call the emergence of this organization lane formation.
We further observe that lane formation can be reversibly tuned by changing the rotation speed.
Details of the reversible dynamics are provided in Appendix~\ref{app:reversible}.

At an intermediate angular velocity $\Omega=\pi/15$~\SI{}{\radian\per\second}, we observe the coexistence of lane formation and a random particle distribution [Fig.~\ref{fig:figure_2}(b)].
The transition between these two states is spatially separated by a characteristic radius $\rho_c\simeq29.2$~\SI{}{\milli\meter}, which is defined below and indicated by the dotted circle in Fig.~\ref{fig:figure_2}(b).
Inside this radius, particles remain randomly distributed even after $\Theta=10$~\SI{}{\radian}, whereas outside the radius, lane formation is already observed at $\Theta=2$~\SI{}{\radian}.

Such a distinct boundary can be understood by considering the particle Weissenberg number, $\mathrm{Wi_p}$, introduced above, defined as
\[
\mathrm{Wi_p} = \frac{\lambda \Omega \rho}{h - 2R}.
\]
Here, the effective gap for the fluid is taken as $h-2R$, rather than the plate separation $h$, because the presence of the particle reduces the available fluid thickness.
When $\mathrm{Wi_p}>1$, the viscoelastic fluid exhibits an elastic response in the vicinity of the particle.
For a given angular velocity $\Omega$, the condition $\mathrm{Wi_p}=1$ defines a characteristic radius
$\rho_c=(h-2R)/(\lambda\Omega)$.
This radius is shown as a red circle in Fig.~\ref{fig:figure_2}
As $\Omega$ increases, $\rho_c$ decreases accordingly. For $\Omega=2\pi/15$~\SI{}{\radian\per\second}, we obtain $\rho_c\simeq14.6$~\SI{}{\milli\meter}.
In this case, the boundary lies close to the shadow of the mechanical connector to the motor, preventing the observation of particles within the red circle.
For $\Omega=\pi/60$~\SI{}{\radian\per\second}, we obtain $\rho_c\simeq116.8$~\SI{}{\milli\meter}. In this case, the boundary is located outside the bottom plate, indicating that the viscoelastic fluid behaves as a viscous fluid throughout the entire region.

We note that $\mathrm{Wi_f}=(\lambda\Omega\rho)/h$ is the Weissenberg number defined in the absence of dispersed particles. In the present experiments, the angular velocity $\Omega$ is chosen such that $\mathrm{Wi_f}<1$ is satisfied throughout the system. As a result, the viscoelastic fluid behaves essentially as a viscous liquid in regions far from the particles. When $\mathrm{Wi_f}>1$, the entire fluid exhibits a dominant elastic response, leading to the onset of bulk fracture~\cite{Tabuteau2009} and preventing stable observation of lane formation.

Closer inspection reveals that the spacing between lanes in the radial ($\rho$) direction is irregular, and that the chain-like structures do not form closed circular rings.
To quantify the lane structure, we computed the normalized spatial autocorrelation of particle positions.
The autocorrelation was evaluated in a rotated reference frame centered on a reference particle aligned with its instantaneous velocity.
A schematic of the procedure is shown in Figs.~\ref{fig:figure_3}(a)--(c).
Details of the construction of the rotated reference frame and the correlation function are described in the following. All particle positions are first expressed in a reference frame centered at the rotation axis of the plates, as shown in Fig.~\ref{fig:figure_3}(a).
For definitions of variables, see Appendix~\ref{app:variables}.
For each particle $i$, we compute its instantaneous velocity $\bm{v}_i=(v_i^x,v_i^y)$ and define the velocity angle $\varphi_i$ by
$|\bm{v}_i|(\cos\varphi_i,\sin\varphi_i)=(v_i^x,v_i^y)$ [Fig.~\ref{fig:figure_3}(b)].
The coordinate frame is centered on $i$-th particle and then rotated by $-\varphi_i$ so that the velocity of the $i$-th particle is aligned along a fixed direction, and spatial correlations are evaluated for all particle pairs $(i,j)$ in this rotated reference frame [Fig.~\ref{fig:figure_3}(c)].

To quantify the spatial organization of particles within a lane, we evaluated the pair correlation function of particle positions in a discretized lattice.
The relative position space is partitioned into square bins
$\bm{\Delta}_{\alpha\beta}=[\alpha\Delta,(\alpha+1)\Delta]\times[\beta\Delta,(\beta+1)\Delta]$,
where $\Delta$ is the lattice spacing and $(\alpha,\beta)\in[-\ell,\ell]\times[-\ell,\ell]$ denote the lattice indices.

For each frame at time $t_k$, we count the number of particle pairs whose relative positions fall within each bin,
\begin{align}
I(\bm{\Delta}_{\alpha\beta},t_k)
= \sum_{i=1}^{n}\sum_{\substack{j=1 \\ j\neq i}}^{n}
\chi\!\left[\bm{r}_{ij}(t_k)\in\bm{\Delta}_{\alpha\beta}\right],
\end{align}
where $n$ is the total number of particles.
Here, $\chi[\cdot]$ denotes the indicator function, which takes the value 1 when the condition in the brackets is satisfied and 0 otherwise.
The relative position $\bm{r}_{ij}(t_k)$ is defined in the rotated frame as
\begin{align}
\bm{r}_{ij}(t_k)
= U(-\varphi_i)\!\left[\bm{r}_j(t_k)-\bm{r}_i(t_k)\right],
\end{align}
with $U(\cdot)$ denoting the rotation matrix.

After averaging over all frames and normalizing by the total number of particle pairs and the bin area, we obtain the probability density
\begin{align}
P(\bm{\Delta}_{\alpha\beta})
= \frac{1}{n^2\Delta^2 N}\sum_{k=1}^{N} I(\bm{\Delta}_{\alpha\beta},t_k),
\end{align}
where $N$ is the total number of analyzed frames.
The resulting distribution $P(\bm{\Delta}_{\alpha\beta})$ is plotted in the $(\Delta_\parallel,\Delta_\perp)$ plane, where $\Delta_\parallel$ and $\Delta_\perp$ correspond to the direction parallel and perpendicular to the particle motion [Figs.~\ref{fig:figure_3}(d)--(f)].

For the low angular velocity $\Omega=\pi/60$~\SI{}{\radian\per\second}, only a weak correlation is observed along the $\Delta_{\parallel}$ direction, appearing as a faint arc-like feature in Fig.~\ref{fig:figure_3}(d), which indicates a largely random particle distribution.
This is further confirmed by the one-dimensional slice of the two-dimensional correlation map along the $\Delta_{\parallel}$ axis [Fig.~\ref{fig:figure_3}(f), blue dashed line].
In contrast, a pronounced correlation along $\Delta_{\parallel}$ emerges for $\Omega=2\pi/15$~\SI{}{\radian\per\second} [Fig.~\ref{fig:figure_3}(e)].
As shown by the orange solid line in Fig.~\ref{fig:figure_3}(f), the correlation function exhibits four distinct peaks at $\Delta_{\parallel}=\pm1$ and $\pm2$~\SI{}{\milli\meter}, indicating a densely packed arrangement of particles along the flow direction.
At the same time, the interparticle distances within these structures exhibit significant fluctuations, characteristic of a liquid-like ordering.
We therefore refer to these elongated, flow-aligned dense regions as chain-like structures.

%====================================
%====================================
\begin{figure}[tb]
\centering
\includegraphics[width=0.48\textwidth]
{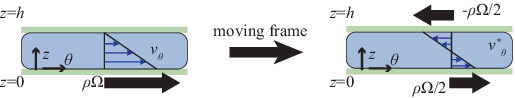}
\caption{\label{fig:figure_4} 
Coordinate transformation to a shifted reference frame used to extract the imposed simple shear flow.
}
\end{figure}
%====================================
%====================================

\subsection{Experiment with a large number of particles: Particle-scale dynamics in chain-like structures}

To focus on the motion of individual particles relative to the imposed shear flow, we introduce a shifted frame with angular velocity $\Omega/2$, corresponding to the average angular velocity of the fixed upper plate and the rotating bottom plate [Fig.~\ref{fig:figure_4}].
In this frame, the azimuthal coordinate of the $i$-th particle is defined as $\theta_i^*=\theta_i-\Omega t/2$.

The angular velocity and azimuthal velocity of the particle in the shifted frame are given by
$\omega_i^*=\omega_i-\Omega/2$ and $v_i^{\theta^*}=\rho_i\omega_i^*$, respectively.
The radial coordinate $\rho$ and the vertical coordinate $z$ are identical to those in the laboratory cylindrical coordinate system.
Thus, the position and velocity of the $i$-th particle at time $t$ are specified by $\rho_i(t)$, $\theta_i^*(t)$, and the two-dimensional velocity vector
$\bm{v}_i^*(t)=(v_i^{\rho}(t),v_i^{\theta^*}(t))$.
See Appendix~\ref{app:variables} for definitions of variables and notation.
In this rotating shifted, a particle located at the mid-plane between the plates in an ideal simple shear flow would have $\omega_i^*=0$.

%==================
%====================================
\begin{figure}[htb]
\centering
\includegraphics[width=0.44\textwidth]{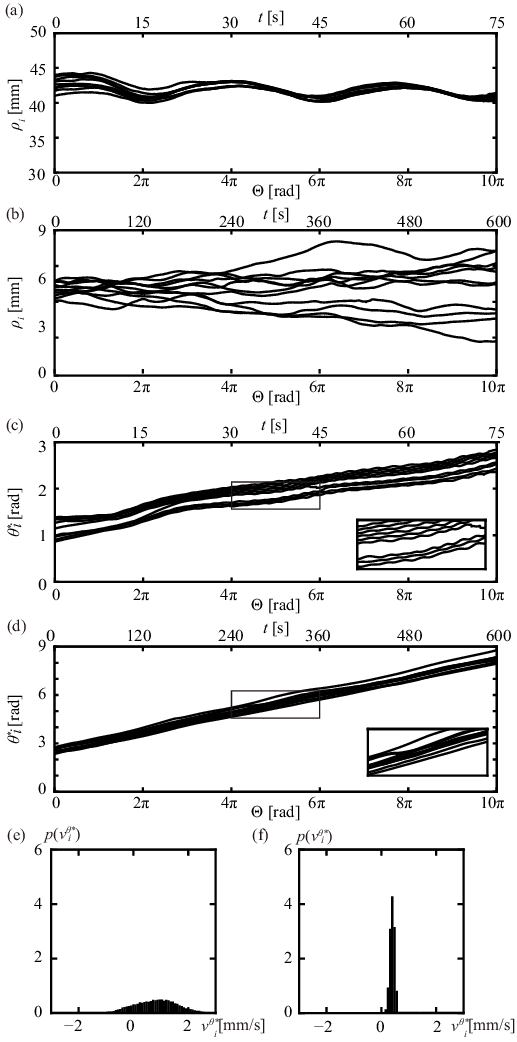}
\caption{\label{fig:figure_5}
Time evolution of individual particle positions in the shifted frame. (a, b) Radial positions $\rho_i(t)$ and (c, d) azimuthal positions $\theta_i^*(t)$ for particles initially located at $\rho_i=40$–$45$~\SI{}{\milli\meter}. Shuttling motion is observed only for $\mathrm{Wi_p}>1$. (e, f) Probability density function $p(v_i^{\theta^*})$ of the azimuthal particle velocity $v_i^{\theta^*}$, evaluated for the same set of particles as in panels (a–d). Here, $\omega_i^*=\omega_i-\Omega/2$ and $v_i^{\theta^*}=\rho_i\omega_i^*$. The distributions are normalized such that $\int p(v_i^{\theta^*})\,dv_i^{\theta^*}=1$. Data are taken over the interval $\Theta=0$ to $2\pi$. Panels (a, c, e) correspond to $\mathrm{Wi_p}>1$ ($\Omega=2\pi/15$~\SI{}{\radian\per\second}), while panels (b, d, f) correspond to $\mathrm{Wi_p}<1$ ($\Omega=\pi/60$~\SI{}{\radian\per\second}).
}
\end{figure}
%====================================
%====================================

The temporal variations of the radial position $\rho_i(t)$ are shown in Figs.~\ref{fig:figure_5}(a) and (b).
For $\Omega=2\pi/15$~\SI{}{\radian\per\second}, we selected particles that formed chain-like structures around $t\simeq30$~\SI{}{\second} and were initially located at $\rho_i=40$–$45$~\SI{}{\milli\meter}.
For $\Omega=\pi/60$~\SI{}{\radian\per\second}, particles with initial radial positions in the same range and located close to each other were selected for consistency with the $\mathrm{Wi_p}>1$ case.

Closer inspection reveals small oscillations of $\rho_i$ with a $2\pi$ period in $\Theta$, which are likely due to a slight misalignment of the parallel plates.
In the case of $\Omega=2\pi/15$~\SI{}{\radian\per\second} [Fig.~\ref{fig:figure_5}(a)], particles exhibit a gradual convergence in the radial direction.
By contrast, for $\Omega=\pi/60$~\SI{}{\radian\per\second} [Fig.~\ref{fig:figure_5}(b)], particles show gradual radial spreading.
This behavior may be qualitatively consistent with dipole-like hydrodynamic interactions reported for particles in viscous fluids in the Stokes regime~\cite{Shani2014-bm,Beatus2012-kz}.

Figures~\ref{fig:figure_5}(c) and (d) show the time evolution of the azimuthal positions $\theta_i^*$ for the same group of particles as in Figs.~\ref{fig:figure_5}(a) and (b), corresponding to $\Omega=2\pi/15$ and $\pi/60$~\SI{}{\radian\per\second}, respectively.
In both cases, a gradual drift of $\theta_i^*$ toward the positive $\theta^*$ direction is observed.
This drift indicates that the particles are located closer to the bottom plate than to the mid-plane, likely due to gravity (see Figs.~\ref{fig:figure_1}(d) and \ref{fig:figure_4} for the geometry).

For $\Omega=2\pi/15$~\SI{}{\radian\per\second} [Fig.~\ref{fig:figure_5}(c)], a pronounced shuttling motion in the $\theta^*$ direction emerges after the formation of chain-like structures at $t\simeq30$~\SI{}{\second}, accompanied by the convergence observed in the radial direction.
By contrast, for $\Omega=\pi/60$~\SI{}{\radian\per\second} [Fig.~\ref{fig:figure_5}(d)], no shuttling motion is observed, and $\theta_i^*$ evolves at an approximately constant rate.

To quantify the behavior of the azimuthal particle velocity $v_i^{\theta^*}$, the probability density function $p(v_i^{\theta^*})$ is shown in Figs.~\ref{fig:figure_5}(e) and (f).
For $\Omega=2\pi/15$~\SI{}{\radian\per\second} [Fig.~\ref{fig:figure_5}(e)], the distribution is broad, reflecting the pronounced shuttling motion of particles.
In contrast, for $\Omega=\pi/60$~\SI{}{\radian\per\second} [Fig.~\ref{fig:figure_5}(f)], the distribution exhibits a sharp peak at a constant positive velocity of approximately $0.4$~\SI{}{\milli\meter\per\second}.

%====================================
%====================================
\begin{figure}[tb]
\centering
\includegraphics{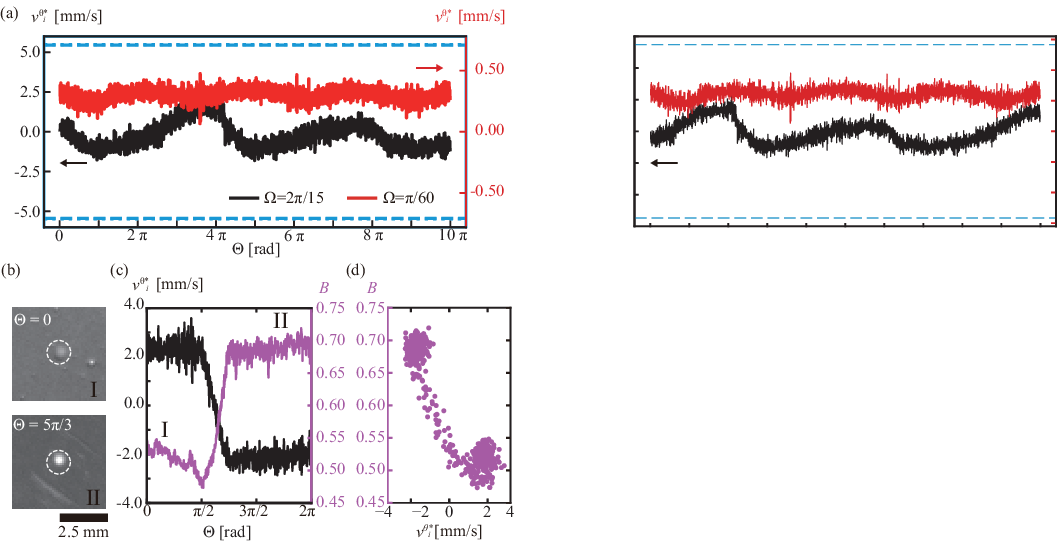}
\caption{\label{fig:figure_6}
(a) Time series of the azimuthal velocity of a single particle, $v_i^{\theta^*}$, measured in a dilute system where interparticle interactions are negligible.
The data shown correspond to particles located at $\rho_i\simeq30$~\SI{}{\milli\meter}.
Black and red lines correspond to $\Omega=2\pi/15$ and $\pi/60$~\SI{}{\radian\per\second}, respectively.
The lower and upper blue dashed lines indicate the velocities of the top and bottom plates at the corresponding radial position $\rho$ and angular velocity $\Omega$, respectively.
(b) Snapshots of a single particle at different values of $\Theta$, showing fluctuations in image brightness in a dyed viscoelastic fluid.
(c) Simultaneous time series of $v_i^{\theta^*}$ (black) and the brightness $B$ (magenta) for a representative particle at $\Omega=2\pi/15$~\SI{}{\radian\per\second}, located at $\rho\simeq38.1$~\SI{}{\milli\meter}.
(d) Scatter plot showing the correlation between $v_i^{\theta^*}$ and $B$, demonstrating a strong negative correlation between particle velocity and brightness.
}
\end{figure}

\subsection{Experiment with a small number of particles: the origin of shuttling motion}

To determine whether the observed shuttling motion arises from interparticle collisions or persists at the single-particle level, we performed experiments with a small number of particles in the absence of particle--particle interactions.
Figure~\ref{fig:figure_6}(a) shows the time series of the azimuthal velocity of an isolated particle over five rotations.
The left and right vertical axes show the azimuthal velocity $v_i^{\theta^*}$ for $\Omega=2\pi/15$ and $\Omega=\pi/60$~\SI{}{\radian\per\second}, respectively.
The data shown correspond to particles located at $\rho_i\simeq30$~\SI{}{\milli\meter}.
The horizontal, lower and upper, blue dashed lines indicate the velocities of the top and bottom plates, respectively, for each case.
In both conditions, $v_i^{\theta^*}$ lies between the two plate velocities. 
For $\Omega=\pi/60$~\SI{}{\radian\per\second}, the fluctuations in $v_i^{\theta^*}$ are small (approximately $0.3$~\SI{}{\milli\meter\per\second}), whereas for $\Omega=2\pi/15$~\SI{}{\radian\per\second} they are significantly larger.

We confirmed that the azimuthal velocity $v_i^{\theta^*}$ for $\Omega=\pi/60$~\SI{}{\radian\per\second} fluctuates around a positive value as in Fig.~\ref{fig:figure_5}(f), implying that the particle is located closer to the bottom plate than to the mid-plane due to gravity. By contrast, $v_i^{\theta^*}$ for $\Omega=2\pi/15$ shows strong fluctuation compared with $\Omega=\pi/60$~\SI{}{\radian\per\second}, indicating the appearance of shuttling motion.% from such positive mean values expected from Fig.~\ref{fig:figure_5}(e).
The large excursions of $v_i^{\theta^*}$ suggest that particles undergo vertical displacements in the gap direction.
In viscoelastic Couette flow, theory predicts that particles migrate toward the moving plates due to the first normal stress difference~\cite{Zhou2020-eu,DAvino2012-ct,Karnis1966-vy}, whereas in purely viscous fluids they tend to accumulate near the mid-plane~\cite{Ho1974-su,Halow1970-ps}.
Assuming an undisturbed linear velocity profile, as shown in Fig.~\ref{fig:figure_4}, such vertical displacements directly modulate the local shear velocity experienced by a particle and hence its azimuthal velocity $v_i^{\theta^*}$.

As a consequence, $v_i^{\theta^*}$ remains nearly constant when $\mathrm{Wi_p}<1$, for which the stable particle height is close to the mid-plane.
By contrast, when $\mathrm{Wi_p}>1$, the particle height becomes bistable, leading to intermittent switching of $v_i^{\theta^*}$ between two distinct values.

To directly test this interpretation, we visualized the vertical motion of particles by dyeing the viscoelastic fluid with 0.02~wt.\% Brilliant Blue FCF.
We rotated the bottom plate from $\Theta=0$ to $2\pi$ at $\Omega=2\pi/15$~\SI{}{\radian\per\second}, corresponding to $\mathrm{Wi_p}>1$, and extracted both the particle velocity and the image brightness $B$ from the recorded images.
As shown in Fig.~\ref{fig:figure_6}(b), $B$ exhibits pronounced temporal fluctuations, reflecting changes in the particle height $z$.

The time series of $B$ and $v_i^{\theta^*}$ are strongly correlated [Figs.~\ref{fig:figure_6}(c) and (d)].
Although small fluctuations in illumination introduce some scatter, $B$ and $v_i^{\theta^*}$ display an approximately linear relationship [Fig.~\ref{fig:figure_6}(d)].
Since the brightness serves as a proxy for the particle height, these results provide direct evidence that vertical position fluctuations are responsible for the observed shuttling of $v_i^{\theta^*}$.

\subsection{Numerical simulation}
%====================================
\begin{figure}[t]
\centering
\includegraphics[width=1.0\columnwidth]{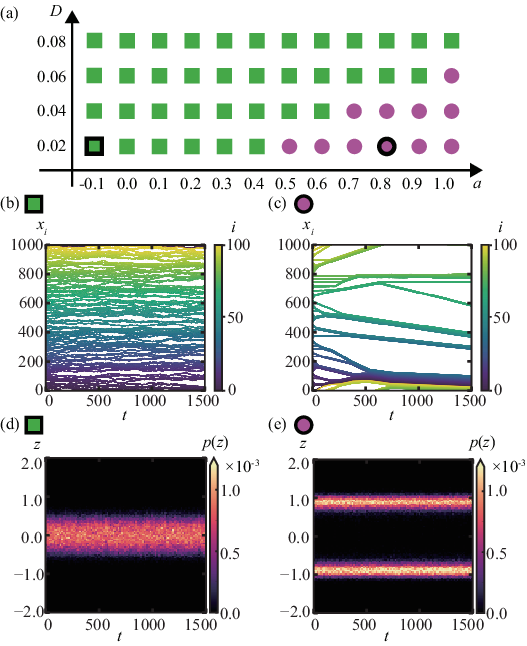}
\caption{\label{fig:figure_7}
(a) Phase diagram of the numerical model [Eqs.~\eqref{eq:model_x}--~\eqref{eq:force}] in the $(a,D)$ parameter space.
Square symbols (green) indicate the random phase, whereas circle symbols (purple) indicate the chain-like phase.
Thick black outlines mark the parameter sets used in panels (b--e).
(b, c) Time evolution of particle positions $x_i$ in the random phase (b) and the chain-like phase (c).
Colors represent the particle index $i$.
(d, e) Time evolution of the probability density function $p(z)$ for the same parameter sets as in (b, c).
The color scale represents the value of $p(z)$, normalized at each time step as $\int p(z)\,dz = 1$.
Parameters are $(D,a)=(0.02,-0.1)$ for (b, d) and $(0.02,0.8)$ for (c, e).}
\end{figure}
%====================================

Experimental results showed that bistability in the particle height $z$ induces intermittent shuttling motion, associated with switching between two stable height levels, in the flow direction when $\mathrm{Wi_p}>1$. This intermittency leads to frequent particle collisions and an effective attraction between particles. To identify the minimal physical mechanism underlying the observed lane formation, we construct a simple mathematical model that explicitly incorporates height bistability and collision-induced interactions.

Specifically, the model is an extension of previously studied quasi-one-dimensional particle models~\cite{Czirok1999-dg,Illien2020-jr}, 
augmented to include a bistable degree of freedom in the height direction, representing viscoelastic migration under strong confinement.

The model describes the overdamped dynamics of $n$ particles in a fluid, resolved in the flow ($x$) and height ($z$) directions.
The system is quasi-one-dimensional, with a finite system size $L$ along the $x$ direction and periodic boundary conditions imposed.
To implement periodic boundary conditions while preserving nearest-neighbor (n.n.) interactions along $x$, 
we define the distance and sign functions as
%%%%%%%%%%%%%%%%%%%%%%%%%%%
\begin{align}\label{eq:defdistSgn}
    \mathrm{dist}(x) &= \mathrm{min}(|x|, |L-x|), \nonumber \\
    \mathrm{Sgn}(x) &= \begin{cases}
     \mathrm{sgn} (x)  & (|x|<|L-x|),\\
    -\mathrm{sgn} (x) & (|x|\geq |L-x|),
  \end{cases}
\end{align}
%%%%%%%%%%%%%%%%%%%%%%%%%%%
where $\mathrm{sgn}(x)=1$ for $x>0$ and $\mathrm{sgn}(x)=-1$ for $x<0$.

The overdamped dynamics of particle motion along the flow ($x$) and height ($z$) directions are described as
\begin{align}
    \Gamma \dot{x}_i &= \frac{V}{h}z_i + \sum_{j \in \mathrm{n.n.}} f_{ij}(x_i), \label{eq:model_x_mod}\\
    \Gamma \dot{z}_i &= a z_i - b z_i^3 + \xi_i(t), \label{eq:model_z_mod}
\end{align}
whose $x_i$ and $z_i$ are the $x$- and $z$- coordinate of $i$ particle, $f_{ij}(x)$ is the interaction force between particles $i$ and $j$, and $\xi(t)$ is the Gaussian white noise  which is defined by $\langle \xi(t) \rangle=0$ and $\langle \xi(t) \xi(t') \rangle= 2D \delta(t-t')$.
The first term in Eq.~\eqref{eq:model_x_mod}, $(V/h)z_i$, represents the coupling between particle height and lateral velocity in a simple shear flow, where $V$ is the plate velocity and $h$ is the gap distance.

Interparticle interactions are modeled by a soft-core repulsive force acting between nearest neighbors,
\begin{align}\label{eq:force_mod}
  &f_{ij}(x_{i})= \nonumber \\
  &\begin{cases}
     k\left\{\mathrm{dist}(x_{i}-x_{j})-2R\right\}\mathrm{Sgn} (x_j-x_i)  & (\mathrm{dist}(x_{i}-x_{j}) \le 2R),\\
    0 & (\mathrm{otherwise}),
  \end{cases}
\end{align}
where $k$ is the spring constant.

As shown experimentally [Fig.~\ref{fig:figure_6}(c)], particles exhibit a bimodal distribution in the height direction.
To capture this behavior, we introduce a Ginzburg--Landau-type double-well potential for the $z$ dynamics.
For $a<0$, the stable height is $z_i=0$, corresponding to the regime $\mathrm{Wi_p}<1$, whereas for $a>0$ the potential becomes bistable with two stable heights at $z_i=\pm\sqrt{a/b}$.
Noise in the $x$ direction is neglected, as the lateral dynamics is dominated by shear advection and interparticle collisions.

Using $R$ and $\Gamma/(bR^2)$ as the units of length and time, respectively (see Appendix~\ref{app:nond} for details), 
the governing equations reduce to
%%%%%%%%%%%%%%%%%%%%%%%%%%%
\begin{align}
    \dot{x}_i &= cz_i + \sum_{j \in \rm n.n.} f_{ij}(x_{i}), \label{eq:model_x}\\
    \dot{z}_i &= a z_i - z_i^3 + \xi_i(t),\label{eq:model_z}
\end{align}
%%%%%%%%%%%%%%%%%%%%%%%%%%%
with the interaction force
\begin{align}\label{eq:force}
  &f_{ij}(x_{i})= \nonumber \\
  &\begin{cases}
     k\left\{\mathrm{dist}(x_{i}-x_{j})-2\right\}\mathrm{Sgn} (x_j-x_i)  & (\mathrm{dist}(x_{i}-x_{j}) \le 2),\\
    0 & (\mathrm{otherwise}),
  \end{cases}
\end{align}
and Gaussian white noise satisfying $\langle \xi(t) \rangle=0$ and $\langle \xi(t)\xi(t') \rangle= 2D\,\delta(t-t')$.

As an initial condition, particles are placed at equal intervals along the $x$ direction, while their $z$ coordinates are drawn from a Gaussian distribution in the range approximately from $-1$ to $+1$.
The equations~\eqref{eq:model_x} and \eqref{eq:model_z} are integrated using the Euler--Maruyama scheme with a time step $\Delta t=0.01$.
We use a spring constant $k=100$, which is sufficiently large to prevent particle overlap while maintaining numerical stability.
The system consists of $n=100$ particles in a domain of size $L=1000$, for which finite-size effects are negligible.
The coupling constant $c$ is set to unity.
The control parameters are varied in the ranges $a\in[-0.1,\,1.0]$ and $D\in[0.02,\,0.08]$.

The phase diagram of the model in the $(a,D)$ parameter space is shown in Fig.~\ref{fig:figure_7}(a).
In this diagram, the random phase is indicated by square symbols ($\square$), whereas the chain-like phase is denoted by circle symbols ($\bigcirc$).
Representative dynamics in the random and chain-like phases are shown in Figs.~\ref{fig:figure_7}(b) and (c), respectively.
Figures~\ref{fig:figure_7}(b) and (c) display the time evolution of particle positions along the $x$ direction, which corresponds to the azimuthal direction in the experiment.

As seen in the phase diagram, the system preferentially aggregates into chain-like structures for small noise intensity $D$ and large values of $a$.
The parameter $a$ controls the effective potential in the $z$ direction; as $a$ is increased, two distinct stable states emerge in $z$, as shown in Figs.~\ref{fig:figure_7}(d) and (e).
These figures display the time evolution of the probability density $p(z)$ for each phase.
In the chain-like phase, $p(z)$ clearly splits into two peaks, indicating bistability in the height direction, whereas in the random phase the distribution remains unimodal and confined within $-0.5 \lesssim z \lesssim 0.5$.
The behaviors in the chain-like phase are consistent with the experimentally observed height bistability [Fig.~\ref{fig:figure_6}(c)], and shuttling motion [Fig.~\ref{fig:figure_5}(c)]. 
Note that the aggregation in the azimuthal direction in the numerical simulations corresponds to the appearance of the peaks in the autocorrelation $P(\Delta_\parallel, \Delta_\perp)$ at $\Delta_\perp=0$ shown in Fig.~\ref{fig:figure_3}(f).
Thus, the quasi-one-dimensional model incorporating a bistable $z$ degree of freedom successfully captures the essential mechanism underlying the chain-like structure.

Previous studies on flow-induced self-assembly have mainly focused on open-channel systems or parallel-plate geometries with a gap much larger than the particle diameter ($h \gg 2R$)~\cite{Michele1977,Pasquino2013}.
By contrast, in our experiments, the gap is comparable to the particle diameter ($h \approx 2R$), leading to strong geometric confinement.
Under this condition, frequent interparticle collisions become unavoidable and play a central role in the formation of chain-like structures, in contrast to earlier studies where hydrodynamic interactions dominate.

\section{Conclusion}

In this study, we developed an experimental system consisting of two parallel rotating acrylic disks with a gap comparable to the particle diameter, which enables in situ observation of particle dynamics under confined shear.
Using this setup, we observed the formation of flow-aligned chain-like structures in a viscoelastic suspension.
Particles form chains when the local Weissenberg number near the particles exceeds unity,
$\mathrm{Wi_p}=\lambda\,\Omega\rho/(h-2R)>1$, whereas they remain randomly distributed for $\mathrm{Wi_p}<1$.

In the $\mathrm{Wi_p}>1$ regime, chain formation is accompanied by a characteristic shuttling motion, manifested as large fluctuations in the azimuthal particle velocity $v_i^{\theta^*}$.
By contrast, for $\mathrm{Wi_p}<1$, the azimuthal velocity remains nearly constant.
Single-particle measurements using a dyed viscoelastic fluid reveal a strong correlation between particle brightness, serving as a proxy for the vertical position, and $v_i^{\theta^*}$.
These observations demonstrate that the shuttling motion originates from a bistability in the particle height $z$, induced by the first normal stress difference, in agreement with previous studies of viscoelastic particle migration~\cite{Michele1977, VanLoon2014}.

To rationalize these findings, we introduced a quasi-one-dimensional numerical model in which the bistable height dynamics, governed by a Ginzburg--Landau-type equation, is coupled to particle translation with excluded-volume interactions.
The resulting phase diagram reproduces the experimental trends: larger bistability parameter $a$ and lower noise intensity $D$ favor chain formation and a bimodal height distribution, while smaller $a$ and larger $D$ lead to a homogeneous random state.
Notably, the intermittency in particle speed induced by height bistability generates an effective attraction analogous to motility-induced phase separation, providing a minimal mechanism for the emergence of chain-like structures.

Taken together, we establish the onset condition for shear-induced alignment in confined viscoelastic suspensions ($\mathrm{Wi_p}>1$), identify the spatial boundary $\rho_c=(h-2R)/(\lambda\Omega)$, and clarify the role of height bistability in generating shuttling dynamics and effective interparticle attraction.
These findings open routes for actively controlling particle organization by tuning the rotation speed $\Omega$ and the confinement gap $h$, and provide a framework for exploring flow-induced self-organization in complex fluids.

\begin{acknowledgments}
This work was supported by JSPS KAKENHI Grant JP16H06478, JP21H00409, and JP21H01004. This work was also supported by JSPS and PAN under the Japan-Poland Research Cooperative Program ``Spatio-temporal patterns of elements driven by self-generated, geometrically constrained flows'', the JSPS Core-to-Core Program ``Advanced core-to-core network for the physics of self-organizing active matter'' (JPJSCCA20230002), and the Cooperative Research of ``Network Joint Research Center for Materials and Devices'' with Hokkaido No.~20181048). 
This work was also supported by the Public Foundation of Chubu Science and Technology Center. This work was also supported by MEXT Promotion of Distinctive Joint Usage/Research Center Support Program Grant Number JPMXP0724020292.
\end{acknowledgments}

\appendix

\section{Variables}\label{app:variables}
%=================================================================
For the reader's convenience, we summarize the variables and coordinate systems used in the experiments, the autocorrelation analysis, and the numerical simulations.
The variables used in the experiments are listed in Table~\ref{table:variables}.
Variables specific to the autocorrelation analysis of particle positions are summarized in Table~\ref{table:ACvariables}.
Most variables used in the numerical simulations are common to those used in the experiments; for clarity, they are listed separately in Table~\ref{table:Simvariables}.

\begin{table}%[b]
\caption{\label{table:variables} Variables used in experiments}
\begin{tabular}{cc}
\hline
Variable & Description \\
\hline
$R$ & particle radius \\
$\Omega$ & angular velocity of the rotating bottom plate \\
$\sigma$ & shear stress \\
$\dot{\gamma}$ & shear rate \\
$\lambda$ & relaxation time of the sample \\
$G',\,G''$ & storage and loss moduli of the sample \\
$G$ & shear modulus of the sample \\
$\eta$ & viscosity of the sample \\
$\omega$ & angular frequency in oscillatory shear \\
$\bm{v}_i$ & velocity of the $i$-th particle \\
$\omega_i=\dot{\theta}_i$ & angular velocity of the $i$-th particle \\
$\omega_i^*=\dot{\theta}_i^*$ & angular velocity of the $i$-th particle in the shifted frame \\
$\Theta=\Omega t$ & accumulated rotation angle of the bottom plate \\
$\mathrm{Wi}_p$ & Weissenberg number near a particle \\
$\mathrm{Wi}_f$ & Weissenberg number far from a particle \\
$h$ & gap distance between the parallel plates
\end{tabular}
\end{table}

\begin{table}
\caption{\label{table:ACvariables} Variables used in the autocorrelation analysis}
\begin{tabular}{cc}
\hline
Variable & Description \\
\hline
$\varphi_i$ & direction of motion of the $i$-th particle \\
$U(\theta)$ & rotation matrix for an angle $\theta$ \\
$\alpha$ & lattice index in the direction parallel to particle motion \\
$\beta$ & lattice index in the direction perpendicular to particle motion \\
$(\alpha,\beta)$ & lattice indices, $(\alpha,\beta)\in[-\ell,\ell]\times[-\ell,\ell]$ \\
$I$ & total number of particle pairs counted in a lattice cell \\
$P$ & probability density function of the autocorrelation \\
$n$ & total number of particles \\
$N$ & total number of frames \\
$i,j$ & particle indices \\
$k$ & frame index \\
$t_k$ & time corresponding to frame $k$ \\
$\Delta_{\alpha\beta}$ & lattice cell indexed by $(\alpha,\beta)$ \\
$\Delta_{\parallel}$ & direction parallel to particle motion \\
$\Delta_{\perp}$ & direction perpendicular to particle motion
\end{tabular}
\end{table}

\begin{table}
\caption{\label{table:Simvariables} Variables used in the numerical simulation}
\begin{tabular}{cc}
\hline
Variable & Description \\
\hline
$x_i$ & position of the $i$-th particle in the flow ($x$) direction \\
$z_i$ & position of the $i$-th particle in the height ($z$) direction \\
$n$ & total number of particles \\
$N$ & total number of time steps \\
$L$ & system size in the $x$ direction \\
$R$ & particle radius \\
$h$ & gap distance between the parallel plates \\
$V$ & plate velocity \\
$f_{ij}$ & interaction force between the $i$-th and $j$-th particles \\
$k$ & spring constant for the soft-core repulsive interaction \\
$\xi_i$ & Gaussian white noise acting on the $i$-th particle \\
$D$ & noise intensity \\
$a,\,b$ & parameters of the double-well potential \\
$c$ & coupling constant after nondimensionalization \\
$\Delta t$ & time step used in numerical integration
\end{tabular}
\end{table}

The position of the $i$-th particle is expressed in Cartesian coordinates as
\begin{align}
\bm{r}_i = x_i \bm{e}_x + y_i \bm{e}_y + z_i \bm{e}_z ,
\end{align}
where $\bm{e}_x$, $\bm{e}_y$, and $\bm{e}_z$ are unit vectors in the $x$, $y$, and $z$ directions, respectively.

We also introduce cylindrical coordinates for each particle,
\begin{align}
\bm{r}_i = \rho_i \bm{e}_{\rho,i} + z_i \bm{e}_z ,
\end{align}
where $\rho_i=\sqrt{x_i^2+y_i^2}$, $\bm{e}_{\rho,i}=(\cos\theta_i,\sin\theta_i)$, and $\theta_i=\arctan\!\big(y_i/x_i\big)$.
The azimuthal unit vector is defined as $\bm{e}_{\theta,i}=(-\sin\theta_i,\cos\theta_i)$, and the positive $\theta$ direction coincides with the rotational direction of the bottom plate.

The particle velocity is written in Cartesian coordinates as
\begin{align}
\bm{v}_i = v_i^x \bm{e}_x + v_i^y \bm{e}_y + v_i^z \bm{e}_z ,
\end{align}
and in cylindrical coordinates as
\begin{align}
\bm{v}_i = v_i^\rho \bm{e}_{\rho,i} + v_i^\theta \bm{e}_{\theta,i} + v_i^z \bm{e}_z .
\end{align}
The angular velocity is given by $\omega_i=d\theta_i/dt=v_i^\theta/\rho_i$.

In the shifted frame shown in Fig.~\ref{fig:figure_4}, the particle velocity is written as
\begin{align}
\bm{v}_i^* = v_i^{\rho} \bm{e}_{\rho^*,i} + v_i^{\theta^*} \bm{e}_{\theta^*,i} + v_i^z \bm{e}_z ,
\end{align}
where the azimuthal velocity in the shifted frame is defined as
\begin{align}
v_i^{\theta^*} = v_i^{\theta} - \rho_i \Omega/2 .
\end{align}

\section{Reversible behavior of lane formation}\label{app:reversible}
%====================================
%====================================
\begin{figure*}[tb]
\centering
\includegraphics{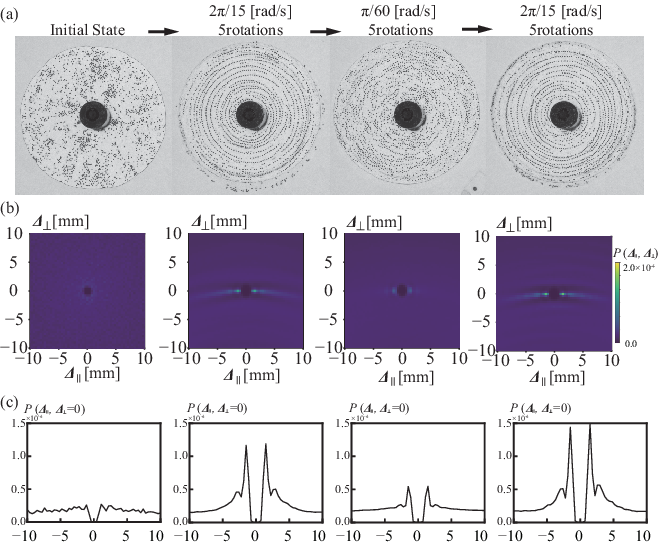}
\caption{\label{fig:figure_sup}
Reversible assembly and disassembly of lane structures composed of chain-like structures by varying the angular velocity of the rotating bottom plate. (a) Snapshots of the particle suspension under sequential changes in the angular velocity $\Omega$. Chain-like structures form when $\Omega = 2\pi/15$~\SI{}{\radian\per\second}, and disappear when $\Omega$ is reduced to $\pi/60$~\SI{}{\radian\per\second}. Such assembly and disassembly can be reversibly controlled by tuning $\Omega$. (b) Spatial autocorrelation function of particle positions, $P(\Delta_\parallel, \Delta_\perp)$, evaluated at each value of $\Omega$. (c) Cross sections of the spatial autocorrelation function $P(\Delta_\parallel, \Delta_\perp)$ at $\Delta_\perp = 0$. These autocorrelation analyses further confirm the reversibility of the assembly and disassembly of chain-like structures.
}
\end{figure*}

Figure~\ref{fig:figure_sup}(a) shows snapshots from experiments in which the angular velocity of the rotating bottom plate $\Omega$ was varied in time. Across the series, once lane formation had developed, reducing the angular velocity to $\Omega=\pi/60$~\SI{}{\radian\per\second} led to the collapse of the previously formed lanes and a chain-like structure.

Figures~\ref{fig:figure_sup}(b) and (c) present the spatial autocorrelation of particle positions, computed in the same manner as Fig.~\ref{fig:figure_3}(d)--(f), respectively. These analyses demonstrate the \emph{reversibility} of lane formation in a Couette shear field: by tuning $\Omega$, an ordered lane state can melt into a disordered configuration and, upon restoring $\Omega$, re-form lanes.%

The dissolution of particle order is reminiscent of the mechanism reported in Poiseuille-like microfluidic flows, where parity-symmetric hydrodynamic dipolar interactions lead to
melt crystalline order~\cite{Saeed2023NP}. In the present system, we observe an analogous ``melting'' under a Couette shear field, indicating that the reversible transition between lane and disordered states can be achieved through controlled changes of the shear rate and confinement.%

\section{Nondimensionalization of variables}\label{app:nond}

The nondimensionalization of the dynamical equations~\eqref{eq:model_x_mod}--\eqref{eq:force_mod} is carried out using
$R$ and $\Gamma/(bR^2)$ as the units of length and time, respectively.
Dimensionless variables and parameters, denoted by tildes, are defined as
\[
\tilde{x}_i=\frac{x_i}{R}, \quad
\tilde{z}_i=\frac{z_i}{R}, \quad
\tilde{t}=\frac{bR^2}{\Gamma}t, \quad
\tilde{a}=\frac{a}{bR^2},
\]
\[
\tilde{\xi}_i=\frac{\xi_i}{bR^3}, \quad
\tilde{D}=\frac{D}{b^2R^6}, \quad
\tilde{k}=\frac{k}{bR^2}, \quad
\tilde{c}=\frac{V}{hbR^2}, \quad
\tilde{L}=\frac{L}{R}.
\]
After nondimensionalization, all tildes are dropped for compact notation.

\end{document}